\documentstyle[prb,aps,epsfig,multicol]{revtex}

\begin{document}
\title{Quantum Critical Behavior and Possible Triplet Superconductivity
in Electron Doped CoO$_2$ Sheets}
\author{D.J. Singh}
\address{Center for Computational Materials Science,
Naval Research Laboratory, Washington, DC 20375}
\date{\today}
\maketitle

\begin{abstract}
Density functional calculations are used to investigate the doping
dependence of the electronic structure and magnetic properties in hexagonal
Na$_x$CoO$_2$. The electronic structure is found to be highly
two dimensional, even without accounting for the structural changes
associated with hydration. At the local spin density approximation level,
a weak itinerant
ferromagnetic state is predicted for all doping levels in the range
$x=0.3$ to $x=0.7$, with competing but weaker itinerant antiferromagnetic
solutions. The Fermi surface, as expected, consists of simple rounded
hexagonal cylinders, with additional small pockets depending on the
$c$ lattice parameter. Comparison with experiment implies substantial
magnetic quantum fluctuations. Based on the Fermi surface size and 
the ferromagnetic tendency of this material,
it is speculated that a triplet superconducting
state analogous to that in Sr$_2$RuO$_4$ may exist here.
\end{abstract}
\begin{multicols}{2}

The last few years have seen the discovery of a number of novel 
unconventional superconductors associated with magnetic phases.
Focusing on triplet (or likely triplet) superconductors these
include
UGe$_2$ (T$_c$ $\sim$ 1K), \cite{saxena}
URhGe (T$_c$ $\sim$ 0.25K), \cite{aoki}
and ZrZn$_2$ (T$_c$ $\sim$ 0.3K), \cite{pfleiderer,zrzn2-note}
where ferromagnetism coexists with superconductivity,
and Sr$_2$RuO$_4$ (T$_c$ $\sim$ 1.5K), \cite{maeno,rice}
which has a paramagnetic Fermi liquid normal state, but is
``near" magnetic phases. Although the exact pairing mechanism
has not been established in these materials, it is presumed 
that spin fluctuations are involved, most probably the quantum critical
fluctuations in the materials with co-existing ferromagnetism
and superconductivity.
\cite{fay,allen,machida,belitz,kirkpatrick,roussev,santi}
In Sr$_2$RuO$_4$, strong nesting related antiferromagnetic
spin fluctuations are found in local density approximation (LDA)
calculations and experiment. \cite{maz-99,sidis}
In addition ferromagnetic fluctuations may
also be present, and if so these would favor a triplet superconducting
state. \cite{maz-97}

Recently, Takada and co-workers
reported the discovery of superconductivity with T$_c$ $\sim$ 5K,
in the CoO$_2$ layer material, Na$_x$CoO$_2\cdot y$H$_2$O. \cite{takada}
Based on the two dimensional 3$d$ transition metal oxide structural motif,
they speculate that the superconductivity may be related to that
of the cuprate high-T$_c$ superconductors. Here an alternate possibility
is discussed, that is the connection with the triplet superconductors
mentioned above. The parent compound of Na$_x$CoO$_2\cdot y$H$_2$O
is Na$_{0.5}$CoO$_2$, which was synthesized and studied in single crystal
form by Terasaki and co-workers. \cite{terasaki,terasaki2}
The compound consists of CoO$_2$ layers (Co is in
at the center of distorted O octahedra formed by layers of O above
and below the Co planes) separated by layers of of Na ions. \cite{hoppe}
The result is a highly anisotropic metal with a large thermopower
at room temperature and a substantial renormalization of the
Fermi liquid properties at low temperature. \cite{terasaki,terasaki2,ando}
Modifications of this compound consisting of substitutions for the
Na layer have been made to obtain technologically useful thermoelectric
materials,
\cite{li,masset,funahashi,miyazaki}
and, in fact, Na$_{0.5}$CoO$_2$ is itself a good thermoelectric
at high temperature. \cite{fujita}
Significantly, the Seebeck coefficient of Na$_{0.5}$CoO$_2$ as calculated
from the LDA band structure using standard kinetic transport
theory is in quantitative agreement with experiment, indicating that
the Fermi liquid description at higher temperature is robust.
\cite{singh-nac}
Band structure calculations for the higher performance
so-called misfit compounds formed by replacement of the Na layer
also reasonably account for room temperature experimental data.
\cite{asahi}

Turning to the low temperature properties, as mentioned, there are
substantial renormalizations in Na$_{0.5}$CoO$_2$,
for example in specific heat and susceptibility.
LDA calculations indicate a ferromagnetic ground state for
this parent compound, while experimentally ferromagnetism is not
observed, one possible explanation being that quantum
fluctuations suppress the ordered ferromagnetic state. \cite{singh-nac}
Interestingly,
the related misfit compound, Ca$_3$Co$_4$O$_9$ shows a large
negative magnetoresistance at low temperature reaching 35\%
at 7T, which could be interpreted as a sign of proximity to a ferromagnetic
quantum critical point.
Other evidence for a proximity to magnetism is the formation of a spin
density wave when Cu doped, \cite{terasaki2} and
the production of a weak magnetic ground state with
strong magnetoresistance upon increasing the Na
concentration to $x=0.75$. \cite{motohashi}

The reported crystal structure of superconducting Na$_x$CoO$_2\cdot y$H$_2$O
differs from that of the parent compound by a reduction in the Na content to
$x \sim 0.35$ and the intercalation of water molecules around the Na sites.
The structure of the CoO$_2$ sheets is very similar to that of the parent
compound Na$_{0.5}$CoO$_2$ -- the in-plane $a$ lattice parameter is slightly
lower (2.823 \AA ~ vs. 2.84 \AA) as is the height of the O (0.885 \AA ~ vs.
an LDA value of 0.908 \AA ~ in the parent). It should be noted that a slight
contraction of the CoO$_6$ octahedra,
as implied by these structural differences,
is consistent with the lower Na content,
in particular a lower electron count on
the CoO$_2$ sheets would be expected to lead to a small contraction.

Here, well converged LDA calculations are reported for Na$_x$Co$_2$O$_4$ for
$x$=0.3,0.5,0.7. In addition, calculations are reported for a strained
lattice corresponding to the structure reported for superconducting
Na$_x$CoO$_2\cdot y$H$_2$O, but neglecting the intercalating water.
The calculations were done using the
general potential linearized augmented planewave method with local
orbitals, \cite{singhbook,singh-lo}
as described in Ref. \onlinecite{singh-nac},
except that better Brillouin zone samples, corresponding to a 
minimum special {\bf k}-points mesh of $16 \times 16 \times 2$
in the hexagonal zone is used here to obtain convergence of the magnetic
energies. As discussed in Ref. \onlinecite{singh-nac}, a virtual crystal
method is used to account for the partially occupied Na site.

Although there is some hybridization,
the valence band structure of Na$_{0.5}$CoO$_2$ consists
of three manifolds of bands separated by gaps -- a lower lying
occupied O 2$p$ derived manifold, followed by Co $t_{2g}$ and $e_g$
manifolds. \cite{singh-nac} As expected from ionic considerations
the Fermi energy, $E_F$ lies near the top of the $t_{2g}$ manifold, which
contains 0.5 holes per Co ion. Because of the actual axial site
symmetry, the $t_{2g}$ manifold can be regarded as consisting of two-fold
(also labeled $e_g$) and one-fold (labeled $a_g$) crystal field
states. These overlap, but the top of the $t_{2g}$ manifold is primarily
of $a_g$ character in Na$_{0.5}$CoO$_2$,
with the result that the band structure near $E_F$ can be
roughly viewed as consisting of one band per Co ion with an filling
of 3/4 (1/2 hole per Co). The present LDA calculations for the
electronic structure of Na$_{0.7}$CoO$_2$ and Na$_{0.3}$CoO$_2$ follow
this picture.
These were done holding the crystal structure fixed at
that of Na$_{0.5}$CoO$_2$ and varying the Na occupancy in the virtual crystal.
The dominant $a_g$ character at the top of the $t_{2g}$
manifold is, however, lost for the strained lattice, discussed below.

\begin{figure}[tbp]
\centerline{\epsfig{file=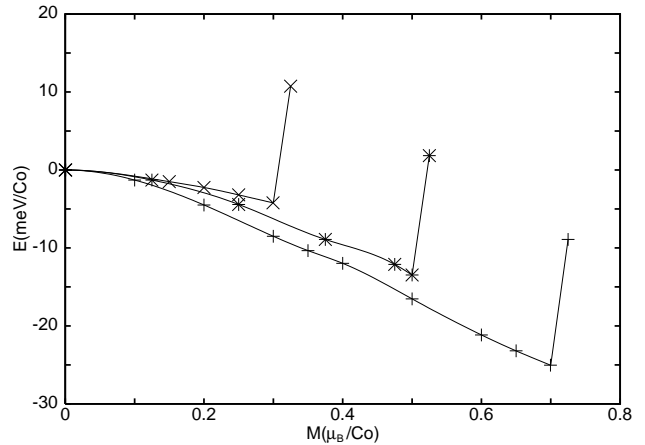,angle=270,width=0.95\linewidth,clip=}}
\vspace{0.2cm}
\nopagebreak
\caption{LDA fixed spin moment energy as a function
of constrained spin magnetization of Na$_x$CoO$_2$ on a per
Co atom basis for $x$=0.7 ($\times$), $x$=0.5 (*) and $x$=0.3
(+). The curves are spline interpolations as a guide to the eye.
Note the breaks at 0.3, 0.5 and 0.7 $\mu_B$ for $x$=0.7, 0.5 and 0.3,
respectively.
These correspond to the band gap between crystal field split Co
$d$ manifolds. The calculations were done keeping the structure
fixed and varying the Na site occupation via the virtual crystal
method (see text).}
\label{fsm}
\end{figure}

The magnetic properties are similar for
the various Na contents in the range $x=0.3$ to $x=0.7$.
Results of fixed spin moment constrained LDA calculations are shown
in Fig. \ref{fsm}. In particular, itinerant ferromagnetism is
found. In each case, the energy decreases with magnetization until
a magnetization at which the band edge is reached in the majority spin.
Then the energy increases rapidly reflecting the crystal field induced
gap between the $t_{2g}$ and $e_g$ manifolds.

Thus, independent of $x$
in this range, the LDA predicts a ferromagnetic ground state, with a spin
moment per Co equal to the number of holes ($p=1-x$) and a half metallic
band structure (here we refer to the hole concentration as
the concentration of holes in the $t_{2g}$ manifold; without
any Na $p$=1; Na electron dopes the sheets, which leads to a reduction
in $p$).
The fixed spin moment curves show a shape crossing
over from parabolic at low moment to more linear as the band edge is 
approached. As may be seen,
the shapes and initial curvatures for the different
doping levels are roughly similar.
The trend towards slightly weaker initial curvatures
at higher Na concentration is possibly an artifact due to the fixed
crystal structure used in the present calculations. It reflects
increasing hybridization (increasing band width and decreasing density
of states) as charge is added to the CoO$_2$ planes. In reality, the
lattice would be expected to expand, perhaps compensating this trend.
In any case, for this range of $x$, ferromagnetism with a magnetic energy
of approximately, $E(FM) \approx 50 p$ in meV/Co and
spin moment $M(FM)=p$ in $\mu_B$/
Co is found. As mentioned, calculations were also done for
a strained cell with the structure of the superconducting sample, but
without H$_2$O. These calculations were done for $x$=0.5 and $x$=0.35,
the latter corresponding to the experimentally determined doping level.
In both cases, the LDA predicted a ferromagnetic state. The magnetic
energies were $E(FM)=$20 and 27 meV/Co for $x$=0.5 and $x$=0.35, respectively.
Thus the behavior is similar, but the magnetic energies are somewhat
larger.

LDA calculations were also done for an antiferromagnetic configuration
with the unit cell doubled along one of the in-plane lattice vectors.
Thus, within a Co plane, each Co ion has four opposite spin
nearest neighbors and two like spin nearest neighbors. At all three
doping levels investigated an antiferromagnetic instability was found,
but this instability is weaker than the ferromagnetic one. Details
of the LDA moments and energies are given in Table \ref{table1}.
Essentially, the energy of the antiferromagnetic configuration examined
tracks the ferromagnetic energy at a value $\sim$ 1/4 as large.

The LDA generally provides a good description of itinerant ferromagnetic
materials. It is known to fail for strongly correlated oxides where
on-site Coulomb (Hubbard) repulsions play an important role in the physics. In
such cases, the LDA underestimates the tendency of the material
towards local moment formation and magnetism. Here, the LDA is found to 
predict
ferromagnetic ground states for materials that are paramagnetic metals
in experiment.
While materials in which the LDA substantially overestimates the tendency
towards magnetism are rare, a number of such cases have been recently
found. These are generically materials that are close to quantum
critical points, and include Sc$_3$In, \cite{aguayo} ZrZn$_2$,
\cite{singh-zrzn2}
and Sr$_3$Ru$_2$O$_7$ (Ref. \onlinecite{singh-327}).
Sr$_3$Ru$_2$O$_7$ displays a novel metamagnetic quantum critical point,
\cite{grigera}
while, as mentioned,
ZrZn$_2$ shows coexistence of ferromagnetism and superconductivity.

Density functional theory is in principle an exact ground state
theory. It should, therefore, correctly describe the spin density
of magnetic systems. However, common approximations to the exact density
functional theory, such
as the LDA,
neglect Hubbard correlations beyond the
mean field level, yielding the underestimated magnetic
tendency of strongly Hubbard correlated systems.
Overestimates of magnetic tendencies, especially in the LDA
are very much less common.
Another type of correlations that is missed in these approximations
are quantum spin fluctuations. This is because the LDA is
parameterized based on electron gases with densities typical for atoms and
solids. However, the uniform electron gas is very far from magnetism in
this density range.
In solids near quantum critical points,
the result is an overestimate of the magnetic moments and tendency
toward magnetism ({\it i.e.} misplacement of the
position of the critical point) due to neglect
of the quantum critical fluctuations.
\cite{yamada,millis}

The present results for Na$_x$CoO$_2$ show a weak ferromagnetic
instability that is robust with respect to doping and structure
(note the instability for the strained lattice).
Based on this, and the experimentally observed renormalized
paramagnetic state, it seems likely that Na$_x$CoO$_2$ is
subject to strong ferromagnetic quantum fluctuations of this type,
and that these are the reason for the disagreement between the
LDA and experimental ground states.

The effects of such quantum fluctuations can be described on a
phenomenalogical level using a Ginzburg-Landau theory
in which the magnetic properties defined by the LDA fixed spin
moment curve are renormalized by averaging with an assumed
(usually Gaussian) function describing the beyond LDA critical
fluctuations. \cite{shimizu} Although a quantitative theory
allowing extraction of this function from first principles calculations
has yet to be established, one can make an estimate based
on the LDA fixed spin moment curves as compared with experiment.
In particular, Na$_x$CoO$_2$ shows a disagreement between the LDA
moment and experiment equal to $p=1-x$, and has a very steeply
rising LDA energy for moments larger than $p$. Thus one may estimate
an r.m.s. amplitude of the quantum fluctuations of $\xi \approx \alpha p$ in
$\mu_B$,
with $ 1/2 < \alpha < 1$, and most likely closer to 1. These are large
values {\em c.f.} ZrZn$_2$.
It is therefore tempting to associate the superconductivity
of Na$_x$CoO$_2\cdot y$H$_2$O with ferromagnetic quantum critical
fluctuations. Considering the simple 2D Fermi surface, which consists
of rounded hexagonal cylinders plus small sections, \cite{singh-nac}
and the ferromagnetic fluctuations a triplet state like that originally
discussed for Sr$_2$RuO$_4$ (Ref. \onlinecite{rice}) seems plausible.
I now speculate about the ingredients in a spin fluctuation mediated
triplet superconducting state.

Within a spin fluctuation induced pairing approach analogous to that
employed for Sr$_2$RuO$_4$ the key ingredient is the integral over
the Fermi surface of the {\bf k}-dependent susceptibility with a function
of the assumed triplet symmetry, \cite{fay,allen,maz-97}
{\em i.e.} in the
simplest case, $ {\bf k} \cdot {\bf k}' / kk'$. For a Fermi surface
in the shape of a circular cylinder, radius $k_F$, the needed integral
is proportional to
$\int_0^{2\pi} {\rm d} \theta cos(\theta) V(2 k_F sin(\theta/2))$,
where $V(k)$ is the assumed pairing interaction. In any case, for
a smooth variation of the spin fluctuations with $k$ and
a maximum at $k$=0 (ferromagnetic), the integral is
roughly proportional to $k_F$ times the
variation of $V$ from $k=0$ to $k=2k_F$.
This latter variation depends on the detailed shape of $V(k)$, but
may be expected to cross over from being proportional to $k_F^2$ for small
$k_F$ to proportional to $k_F$ for larger $k_F$. Neglecting small Fermi
surface sections, $k_F$ varies as $p^{1/2}$. One possibility for $V(k)$
is a function smoothly going from a finite value at $k=0$ to near zero
at the zone boundary (reflecting the rather weak antiferromagnetic
instability relative to the ferromagnetic), with a size at $k=0$
given by the LDA ferromagnetic energy ($\propto p^2$) or alternately
a Hund's coupling ($p$ independent)
times $\xi$ ($\propto p$).

Within this $p$-wave scenario it would be quite interesting
to measure the variation of the superconducting properties of
Na$_x$CoO$_2\cdot y$H$_2$O as a function of doping level. The above
arguments imply a substantial model dependent variation up to the level
where proximity to the critical point suppresses $T_c$, with the
implication that still higher values of $T_c$ may be obtained.
However, it should be emphasized that the mechanism and superconducting
symmetry of Na$_x$CoO$_2\cdot y$H$_2$O have yet to be established, and
in fact, even conventional electron-phonon superconductivity competing
with spin fluctuations has not been excluded.

I am grateful for helpful discussions with R. Asahi,
I.I. Mazin, A.J. Millis, W.E. Pickett, S.S. Saxena and I. Terasaki.
Some computations were performed using facilities of the DoD HPCMO ASC
and ARL centers. The DoD-AE code was used for some of this work.

\begin{table}
\caption{LSDA spin magnetizations and energies for Na$_x$CoO$_2$.
All quantities are on a per Co basis. Energies are in meV,
spin moments are in $\mu_B$, FM denotes ferromagnetic and AF
denotes the partially frustrated nearest neighbor AF configuration
discussed in the text. M is the total spin magnetization, m is the
magnetization inside the Co LAPW sphere, radius 1.95 Bohr.
}
\begin{tabular}{dddddd}
\hline
     & E(FM) & M(FM) & m(FM) & E(AF) & m(AF)  \\
\hline
x=0.3 & 25. & 0.70 & 0.56 & 9.  & 0.36  \\
x=0.5 & 13. & 0.50 & 0.41 & 3.  & 0.21  \\
x=0.7 &  4. & 0.30 & 0.25 & $\leq$ 1.  & 0.04  \\
\hline
\end{tabular}
\label{table1}
\end{table}

\end{multicols}

\begin{references}
\bibitem{zrzn2-note}
An alternate, Fulde-Ferrel-Larkin-Ovchinikov, non-triplet state is not
excluded in ZrZn$_2$. \cite{singh-zrzn2}
\end{references}
\end{document}